\documentclass[aip, preprint
 amsmath,amssymb,
 reprint,%
]{revtex4-1}

\usepackage{graphicx}
\usepackage{dcolumn}
\usepackage{bm}
\usepackage[utf8]{inputenc}
\usepackage[T1]{fontenc}
\usepackage{mathptmx}
\usepackage{etoolbox}
\usepackage{xcolor}

\makeatletter
\def\@email#1#2{%
 \endgroup
 \patchcmd{\titleblock@produce}
  {\frontmatter@RRAPformat}
  {\frontmatter@RRAPformat{\produce@RRAP{*#1\href{mailto:#2}{#2}}}\frontmatter@RRAPformat}
  {}{}
}%
\makeatother
\begin{document}

\preprint{AIP/123-QED}

\title{Partial immunity of two-photon correlation against wavefront distortion\\for spatially entangled photons}

    \author{Kiran Bajar}
    \author{Rounak Chatterjee}
    \author{Vikas S. Bhat}
    \author{Sushil Mujumdar*}
     \email{mujumdar@tifr.res.in}
    \affiliation{%
     Department of Nuclear and Atomic Physics, Tata Institute of Fundamental Research, 400005 Mumbai, India
    }

\date{\today}

\begin{abstract}
  High-dimensional quantum entanglement in photons offers notable technological advancements over traditional qubit-based systems, including increased information density and enhanced security. However, such high-dimensional states are vulnerable to disruption by complex disordered media, presenting significant challenges in practical applications. Spatially-entangled photons are conventionally generated using a nonlinear crystal via spontaneous parametric down conversion (SPDC). While the effect of disorder on spatially entangled photons in the near field of the crystal is well understood, the impact of disorder in the far field is more complex. In this work, we present a systematic study of the randomization of two-photon correlations caused by arbitrary phase distortions in the far field by breaking it down into odd and even parity components. First, we theoretically show that the two-photon field is only sensitive to the even-parity part of the phase distortion. In follow-up experiments, we employ a deformable mirror to implement random phase distortions, separating the contributions of odd and even parity phases using Zernike polynomials. The experimental results are in agreements with the theoretical predictions. Subsequently, we perform numerical simulations to show that these results extend to stronger degrees of  disorder. Our key finding is that, since two-photon correlations are only affected by the even-parity component of phase modulations, the number of independent adaptive optics elements required for optimizing the correlation can be effectively halved, offering a significant practical advantage in managing disorder in quantum systems.
  
\end{abstract}

\maketitle

\section{\label{sec:level1}Introduction}
High-dimensional entanglement is gaining popularity because of its ability to provide high bandwidth and error tolerance in quantum communication \cite{Cerf2002,HighD}. A commonly used method to generate high-dimensional entangled photons is spontaneous parametric down-conversion (SPDC) \cite{Klyshko2018,SPDC,DiLorenzoPires2011} in nonlinear crystals. Using SPDC, high-dimensional entanglement can be realized across various degrees of freedom, such as position-momentum (spatial) entanglement \cite{PMEnt}, energy-time entanglement \cite{EnTime}, and angle-angular momentum entanglement \cite{Mair2001, Krenn2017}. Of these, spatial entanglement is particularly appealing as spatial modes can be efficiently controlled using adaptive optics \cite{Boucher2021, Lib2024}.

When optical disorder is introduced in the path of high-dimensional spatially entangled photons, no interference is observed in the intensity distribution \cite{2pspeckles}. This occurs because the single-photon state comprises a statistical ensemble of multiple spatial modes and is, therefore, incoherent in nature \cite{Glauber1963}. Instead, the interference is observed in the coincidence space of two photons, which is referred to as two-photon interference \cite{hom,nonco1} and the random interference pattern as two-photon speckles \cite{Beenakker2009,2pspeckles}. The two-photon speckle indicates the randomization of correlation between the two photons. Hence, the scattering of photons poses a significant challenge in various quantum technologies, such as for instance, long-distance high-dimensional quantum key distribution \cite{Etcheverry2013,Otte2020,Zhou2019} and quantum imaging \cite{Unternahrer2018,He2023}, and the rectification of effects of disorder is a crucial field of research \cite{Valencia2020}. To that end, earlier studies have shown that the two-photon interference caused by disorder in the near field of a nonlinear crystal can be controlled using a feedback mechanism \cite{realtime} and a transmission matrix approach \cite{Courme2023}. These interference patterns correlate with those produced by a coherent pump beam of half the wavelength when both entangled photons traverse the disorder \cite{Lib2020}. Additionally, the patterns correlate with the interference produced by a coherent beam of the same wavelength when only one of the two photons passes through the disorder \cite{Devaux2023}. The two-photon correlations can be controlled in real-time using feedback from the coherent beam speckle \cite{realtime}. However, if the disorder is induced in the far field of the crystal, the two-photon wavefunction changes as it propagates to the far field \cite{Chan2007}. The two-photon interference pattern then does not correlate with the coherent beam's interference pattern. In this case, restoring and controlling two-photon interference caused by disorder is more challenging \cite{Black2019}.

 In this work, we address this issue and  present a method to retrieve the two-photon correlation when disorder is introduced in the far field of collinear SPDC process. We examine the relationship between two-photon speckles in relation to the speckles of a coherent auxiliary pump beam. Initially, the two interference patterns appear uncorrelated; however, a parity check on the disorder reveals signature of a correlation. Our theoretical results show that while the auxiliary pump carries information about the entire disorder, the two-photon interference pattern contains information only about the even-parity component of the disorder. The odd-parity component does not affect the two-photon correlation. To demonstrate this experimentally, we use a deformable mirror (DM) as a controllable lossless disorder proxy, and implement even-parity and odd-parity parts of the disorder using different linear combinations of Zernike polynomials programmed into the DM. The experimental results are in excellent agreement with the simulation outcomes, indicating that the two-photon correlation remains unaffected by the odd-parity components of the disorder. This finding significantly simplifies the restoration of two-photon correlation, as it requires correction of only the even-parity components. This fact has substantial implications in quantum communication and imaging, based on the alleviation of technical requirements of wavefront correction methods. The results are applicable in scenarios where both photons traverse the same channel, making them relevant for super-dense coding \cite{Hu2018,Guo2019}, aberration correction in quantum microscopy \cite{Unternahrer2018}, and various protocols of quantum key distribution \cite{Long2002,Cabello2000} and quantum communication \cite{Deng2003}. Additionally, these results can be extended to non-collinear SPDC by appropriately shaping the auxiliary pump beam.

\section{Theoretical framework}
The two-photon wavefunction generated by collinear degenerate SPDC in terms of transverse momentum ($k_1,k_2$) can be expressed as follows \cite{Howell}:
\begin{eqnarray}
\psi(k_1,k_2) \propto e^{-(k_1+k_2)^2\sigma_+^2/{2}}e^{-(k_1-k_2)^2\sigma_-^2/{2}}|k_1\rangle |k_2\rangle
\label{eq:one}
\end{eqnarray}

where $\sigma_+ = w_o/\sqrt{2}$ and $\sigma_- = \sqrt{L \lambda_p/{12\pi n_p}}$. Here, $L$ denotes the length of the crystal, $n_p$ is the refractive index at the pump wavelength $\lambda_p$, and $w_0$ is the waist size of the pump beam.

We introduce a phase modulation, described by the transfer function $A_d(k) = \exp(i\phi(k))$, in the far field of the crystal plane. The wavefunction at the output of this phase modulation is given by:
\begin{eqnarray}
\psi(k_1,k_2) &\propto& A_d(k_1)A_d(k_2)e^{-(k_1+k_2)^2\sigma_+^2/{2}}\nonumber\\
&&e^{-(k_1-k_2)^2\sigma_-^2/{2}} |k_1\rangle |k_2\rangle
\label{eq:two}.
\end{eqnarray}
The dimensionality of spatial entanglement is characterised by the Schmidt number\cite{Law2004,Bhat2024}. In the regime of high Schmidt number, the single photon SPDC wavefunction contains  a large number of spatially incoherent modes, leading to the absence of interference in direct intensity measurements. The interference pattern is revealed in the coincidence detection of the two photons.

Mathematically, the two-photon interference pattern can be derived from the correlation function, $C(x_1,x_2) = \langle a^\dagger(x_1)a^{\dagger}(x_2)a(x_2)a(x_1)\rangle$, where the field operator $a(x) = \int dk e^{-ikx}a(k)$. 
Thus, for the quantum state given above, the two-photon interference pattern is:
\begin{eqnarray}
C(x_1,x_2) \propto& \left|\int\int dk_1 dk_2 e^{ik_1x_1}e^{ik_2x_2} A_d(k_1)A_d(k_2) \right.\nonumber \\
&\left. e^{-(k_1+k_2)^2\sigma_+^2/{2}}e^{-(k_1-k_2)^2\sigma_-^2/{2}} \right|^2\langle 0|0 \rangle \langle 0|0 \rangle
\label{eq:three}
\end{eqnarray}

where $|0\rangle$ represents vacuum state. For a thin crystal, high-dimensional entanglement can be achieved under two conditions: (a) $\sigma_+ \gg \sigma_-$ or (b) $\sigma_- \gg \sigma_+$. Experimentally, the former condition is easier to realize, as it requires a well-collimated pump beam, rather than a tightly focused beam, which diverges rapidly.

Furthermore, considering that the length scale of randomness in the phase modulation in k-space is much larger than $1/\sigma_+$ (i.e., weak disorder), the term $e^{-(k_1 + k_2)^2 \sigma_+^2 / 2}$ can be effectively replaced by $\delta(k_1 + k_2)$ for $\sigma_+ \gg \sigma_-$. Here, $\delta$ denotes the Dirac delta function, implying that the two photons have equal and opposite transverse momenta and thus appear at radially opposite points in the far field plane. Under this approximation, the coincidence correlation function simplifies to:
\begin{eqnarray} 
C(x_1,x_2)\propto \left|\int dk_1 e^{ik_1(x_1-x_2)}A_d(k_1)A_d(-k_1)e^{-2k_1^2\sigma_-^2}\right|^2 \nonumber\\
\end{eqnarray}
which indicates that interference pattern will appear in the coincidence space defined by $x_1 - x_2$ \cite{2pspeckles}.

Substituting the expression for $A_d(k)$ into the equation yields:
\begin{eqnarray}
C(x_1,x_2)\propto \left|\int dk_1 e^{ik_1(x_1-x_2)} 
e^{i(\phi(k_1)+\phi(-k_1))}
e^{-2k_1^2\sigma_-^2}\right|^2 \nonumber\\
\label{eq5}
\end{eqnarray}

This form shows that for phase modulation $\phi(k)$, $\phi(k)+\phi(-k)$ plays a critical role in the interference pattern observed in the two-photon coincidence detection, and leads to the formation of two-photon speckles in the $x_1 - x_2$ space.

\begin{figure*}[t]
\includegraphics[width=\linewidth]{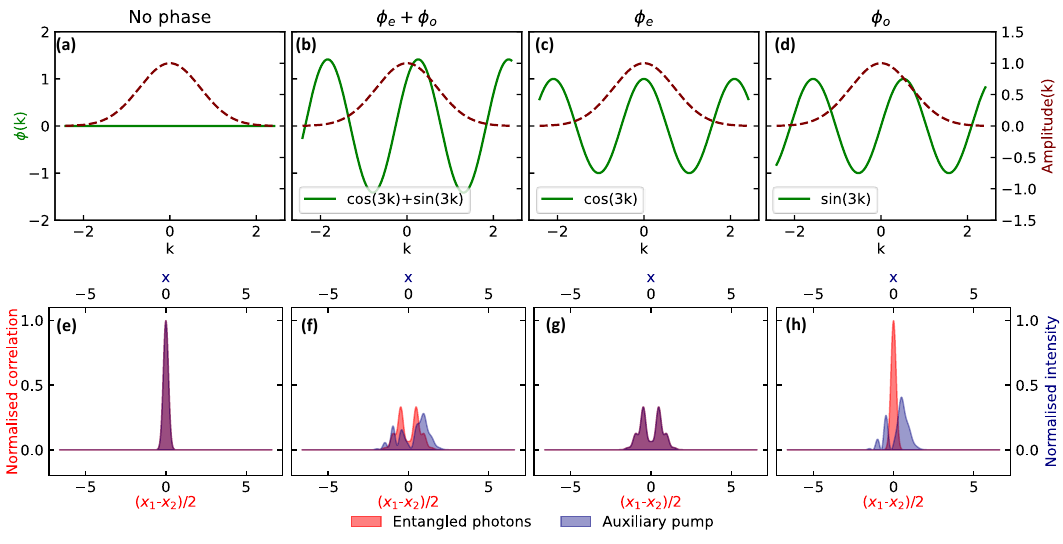}
\caption{\label{fig:1} Numerical calculations of two-photon interference of spatially entangled photons and its comparison with interference pattern of auxiliary pump beam. The calculations are made using Eq.~\ref{eq6} and Eq.~\ref{eq7}. Panels (a)-(d) represent the amplitude of the initial Gaussian beam (red dashed curves) in k-space and the phase modulations (green solid curves), with the phase function mentioned in the respective plot. $\phi_o$ and $\phi_e$ in (c) and (d) are the odd and even components of the phase function in (b). Panels (e)-(h) represent the two-photon interference pattern (red shade) in $(x_1-x_2)/2$ and auxiliary pump intensity interference pattern (blue shade) in x for the same phase modulation. (a) \& (e) correspond to the case when no phase is introduced, where the two interference patterns overlap exactly. (f) Single photon interference and two-photon correlations are affected by the phase. (g) For even phase, the effect on both is identical. Note also that the two-photon correlation patterns matches that in (f). (h) For odd phase, only the single photon interference is affected, while the two-photon correlation is preserved. The intensity and correlation patterns are normalised to maintain a constant area under the curve.}
\end{figure*}

The symmetry of the wavefuntion suggests that we should examine the odd-parity ($\phi_o(k)$) and even-parity ($\phi_e(k)$) components of $\phi(k)$ \cite{Black2019,Bonato2008}, i.e, $\phi_e(k) = (\phi(k)+\phi(-k))/2$ and $\phi_o(k) = (\phi(k)-\phi(-k))/2$. According to Eq.~\ref{eq5}, only the even-parity component of the phase modulation contributes to the two-photon interference. The odd-parity component does not contribute to interference pattern because the equal and opposite phases of the two photons cancel each other out. 

Thus the two-photon interference pattern is finally given by:
\begin{eqnarray}
C(x_1,x_2) \propto& \left|\int dk_1 e^{ik_1(x_1-x_2)}e^{i2\phi_e(k_1)}e^{-2k_1^2\sigma_-^2}\right|^2
\label{eq6}
\end{eqnarray}

Now, consider a coherent Gaussian beam with half the wavelength of the photons and the same beam waist ($\sqrt{2}/\sigma_-$) in  the k-plane. Consider that the beam experiences the even-parity component of the phase modulation, $\phi_e(k)$. 
Since a photon with half the wavelength acquires twice the phase shift, the effective transfer function for the coherent beam is $e^{i2\phi_e(k)}$. Therefore, the intensity interference pattern for this beam is given by:
\begin{eqnarray}
I(x) \propto& \left|\int dk_p e^{ik_px}e^{i2\phi_e(k_p)}e^{-k_p^2\sigma_-^2/2}\right|^2
\label{eq7}
\end{eqnarray}
where the wavenumber $k_p = 2k_1$. It is evident that the two-photon interference pattern described by Eq.~\ref{eq6} has the same functional form as the interference pattern of the coherent beam given by Eq.~\ref{eq7}. Since this coherent beam has the same wavelength as the pump beam used in degenerate SPDC, we refer to it as the `auxiliary pump'. 

To illustrate the theoretical results, we present a simple numerical example. Fig.~\ref{fig:1} shows the calculation for two-photon interference using the phase modulation $\phi(k)=\cos(3k)+\sin(3k)$, and compares it with the interference pattern of the auxiliary pump beam. 
Fig.~\ref{fig:1} (a)-(d) depict the amplitude of the beam in the k-space (red dashed curve, right Y-axis) and the phase introduced onto the beam (green solid curve, left Y-axis). The subplots illustrate (a) when no phase is introduced, (b) $\phi(k)=\cos(3k)+\sin(3k)$ is applied, (c) when only the even-parity component $\phi_e(k)$ is introduced, and (d) when only the odd-parity part $\phi_o(k)$ is introduced. Fig~\ref{fig:1} (e)-(h) display the corresponding two-photon interference patterns (red shade) and auxiliary pump interference pattern (blue shade) for these four cases. These patterns show that the auxiliary pump beam is influenced by both the even-parity and odd-parity components, whereas the entangled photon pair experiences only the even-parity part. Specifically, when the phase is odd-parity (Fig.~\ref{fig:1} (h)), the auxiliary pump exhibits an interference pattern, but the entangled photons show no two-photon interference. Conversely, when an even-parity phase modulation is introduced (Fig.~\ref{fig:1} (g)), the interference pattern of the auxiliary pump matches the two-photon interference pattern in the variable $(x_1-x_2)/2$. Also, the two-photon interference pattern corresponding to the total phase $\phi$ (Fig.~\ref{fig:1} (f)) is identical to the interference pattern observed when only $\phi_e$ is applied (Fig.~\ref{fig:1}(g)). This confirms that only the even-parity component of the phase modulation contributes to the interference of entangled photons.

\section{Experiment and Results}
\subsection{Experimental Setup}
    \begin{figure*}[ht!]
    \includegraphics[width=\linewidth]{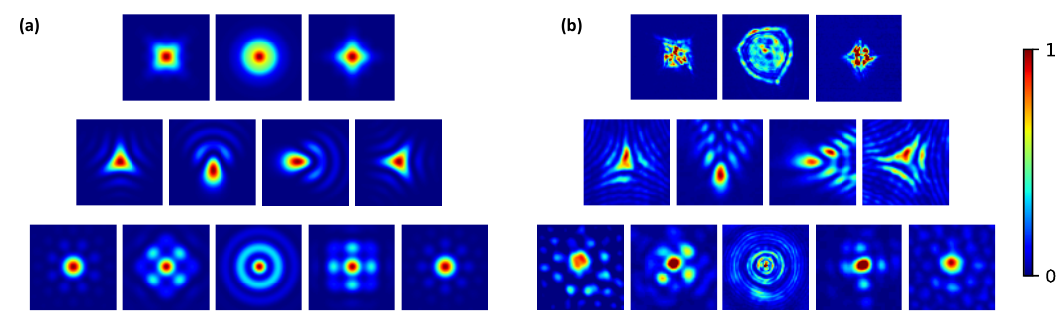}
    \caption{\label{fig:2} (a) Numerically-computed intensity patterns obtained in the far field using Zernike polynomials ($\mathrm{Z_4}$-$\mathrm{Z_{15}}$) when illuminated by a Gaussian beam. (b) Experimentally obtained intensity patterns from the $40$-element DM with like illumination. Deviations from theory arise from finite (and large) pixel size and inherent hardware limitations in the DM. The first row shows $\mathrm{Z_4}$-$\mathrm{Z_6}$ (left to right), the second row shows $\mathrm{Z_{7}}$-$\mathrm{Z_{10}}$ (left to right) and the third row shows $\mathrm{Z_{11}}$-$\mathrm{Z_{15}}$ (left to right).}
    \end{figure*}

    To systematically investigate the effects of the odd and even parity components of phase modulation experimentally, we utilize a deformable mirror (DM) to implement the phase modulation. The DM surface can generate Zernike polynomials, which have well-defined parity and can be controlled independently (see Appendix \ref{ap:DM} for more details on the DM and Zernike polynomials). By applying different combinations of odd and even Zernike modes, we can introduce precise phase modulations to study their distinct effects on two-photon interference patterns. To extract the Zernike modes from the DM, we illuminate it with a coherent Gaussian beam that has a waist radius comparable to the DM’s pupil radius and observe the resulting beam in the far field. Fig.~\ref{fig:2}(a) shows the numerical simulation of this setup, while Fig.~\ref{fig:2}(b) presents the experimental results. The intensity patterns observed experimentally agree with the simulation, confirming the effectiveness of the Zernike mode control. The slight mismatch between the experimental and simulated patterns can be attributed to the known hysteresis effect of the piezoelectric material used in the DM \cite{Tyson2022}.
    
\begin{figure}
\includegraphics[width=\linewidth]{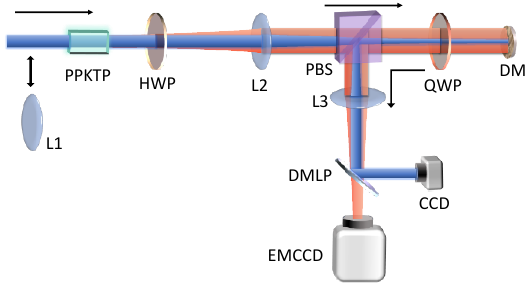}
\caption{\label{fig:3} Experimental setup. L: Lens, PPKTP: Periodically-poled potassium titanyl phosphate crystal, HWP: Half Wave-plate for $810$ nm, PBS: Polarizing beamsplitter, QWP: Quarter Wave-plate for $810$ nm, DM: Deformable mirror, DMLP: Longpass dichroic mirror, EMCCD: Electron multiplying charge coupled device, CCD: Charge coupled device. Lens L1 is inserted before the crystal to observe $405$ nm interference pattern using CCD.}
\end{figure}

    Fig.~\ref{fig:3} illustrates the experimental setup. A vertically polarized pump beam with a wavelength of $405$ nm and a waist size of $\approx 425 \mu$m is directed through a $5$ mm long Type-0 periodically poled potassium titanyl phosphate (PPKTP) crystal. The crystal generates vertically polarized, degenerate photon pairs at a wavelength of $810$ nm through SPDC. The emitted photon pairs exhibit spatial entanglement, with a Schmidt number of $\approx 700$ (see Appendix \ref{ap:SNo}). The far field of the crystal’s output face is imaged onto the DM using lens L2. The crystal is positioned on the front focal plane of L2, and the DM is located on the back focal plane. The focal length of L2 was carefully chosen to ensure the entire area of the DM is covered by photon pairs, maximizing its effective utilization. A polarizing beamsplitter (PBS) is used to separate the incident and reflected beams from the DM. Before L2, a half wave- plate (HWP) rotates the polarization of the $810$ nm photons to horizontal, allowing them to be transmitted through the PBS and directed to the DM. A quarter wave-plate (QWP), placed between the PBS and DM, rotates the polarization back to vertical after reflection. A long-pass dichroic mirror (DMLP) separates the pump beam from the down-converted photons, directing them to separate detection paths. The pump beam is detected using a CCD camera, while the down-converted photons are captured using an electron-multiplying charge-coupled device (EMCCD). The deformations introduced by the DM are observed in the far field using lens L3, with the DM positioned at the front focal plane and the detectors at the back focal plane of L3. To block stray photons, an $810\pm5$ nm bandpass filter (not shown) is mounted in front of the EMCCD.

In the initial setup (Fig.~\ref{fig:3}), the pump beam is collimated and then focused onto the deformable mirror (DM) by lens L2. Since the pump beam is focused onto the center of the DM, it does not undergo any deformations from the DM, and thus no interference pattern is observed. To capture the effects of the DM using the pump beam, we introduce the auxiliary pump by adding a lens combination with effective focal length $3$ cm (represented here as L1) before the crystal. This focuses the beam onto the crystal. Lens L2 then collimates the auxiliary pump beam again and directs it to the DM. The term 'auxiliary pump' is used here to indicate that we are no longer detecting the down-converted photons but instead analyzing the deformations of the pump beam itself, using the CCD camera. We verified that the disorder strength applied by the DM for the two wavelengths, $405$ nm and $810$ nm, is comparable (see Appendix \ref{ap:SC}). The two-photon interference pattern (without L1) from the entangled photons is recorded separately using the EMCCD. These measurements are taken consecutively, and the interference pattern of the auxiliary pump is compared with the two-photon interference pattern.

The focal lengths of lenses L2, and L3 are $40$ cm, and $50$ cm, respectively. To ensure that both the auxiliary pump and the down-converted photons experience the same deformations, two conditions are met: (a) the beams should have the same size on the DM, achieved by selecting the appropriate focal length for L1, and (b) they must overlap perfectly on the DM plane, ensured by carefully aligning L1.

\begin{figure}[b]
\includegraphics[width=\linewidth]{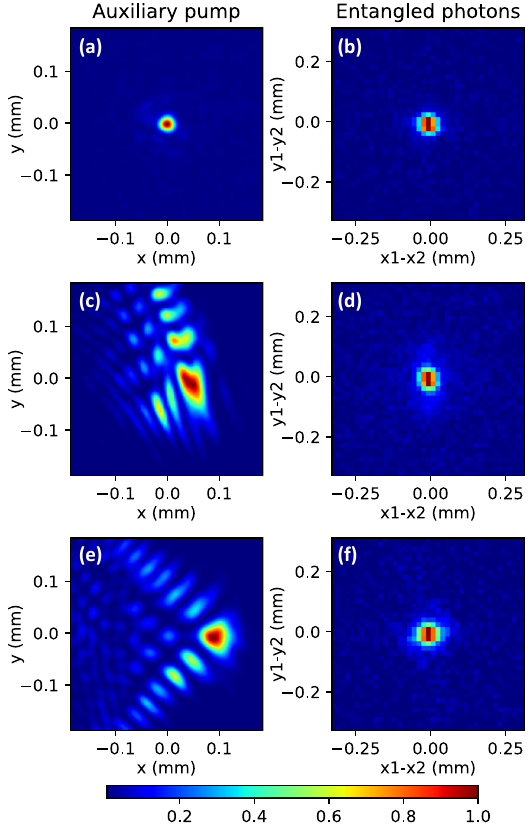}% Here is how to import EPS art
\caption{\label{fig:4} Experimental results for no phase (first row) and two different configurations of odd-parity disorder (second and third row) implemented using DM. Panels (a), (c) and (e) show the auxiliary pump interference pattern and panels (b), (d) and (f) represent two-photon correlation in $x_1-x_2$ coincidence space corresponding to the same configuration. Odd parity disorder has no effect on two-photon correlation.}
\end{figure}

For two-photon correlation measurements, we employed an EMCCD camera as a multi-pixel photon detector, following established techniques for photon counting with EMCCDs \cite{emccd,Hugo,MTh}. The crux of these techniques is as follows. The EMCCD captures 
$N$ frames with very low photon exposure, where the average photon count is approximately $0.1$ per pixel per frame. Each frame records photo-electron counts at each pixel, which are then converted to photon numbers by applying a threshold $T$. This threshold is determined using the mean ($\mu$) and standard deviation ($\sigma$) of the photo-electron counts as: $T = \mu + \sigma$. Pixels receiving more than $T$ number of photo-electrons are considered to have received one photon, while those below $T$ are considered photon-less.

The frames are analyzed post-capture using coincidence detection and background subtraction to extract the two-photon correlation. The 2D correlation function, denoted as 
$Corr(x,y)$, is computed as:
\begin{eqnarray}
\text{Corr}(x,y) &= &\frac{1}{N}\sum_{i=0}^{N-1}\left(\sum_{x_1,y_1} P_i(x_1,y_1)P_i(x_1+x,y_1+y)\right.\nonumber\\
&&\left.- \sum_{x_1,y_1} P_i(x_1,y_1)P_{i+1}(x_1+x,y_1+y)\right)
\label{eq:ten}
\end{eqnarray}

where $Corr(x,y)$ represents the two-photon correlation function as a function of the pixel position difference $(x,y)$ and $P_i(x_1,y_1)$ is the number of photons detected in the $i^{th}$ frame at pixel $(x_1,y_1)$. The first term in summation measures the number of photon coincidences at pixel separations $(x,y)$, while the second term removes accidental coincidences by correlating photons in consecutive frames. 

\subsection{Experimental Results}
Fig.~\ref{fig:4} presents the experimental results comparing the auxiliary pump's interference pattern with the two-photon correlations for various odd-parity configurations of the DM. For each configuration, $50,000$ frames were captured using the EMCCD. The left column of the figure shows the interference patterns of the auxiliary pump, while the right column shows the two-photon correlations corresponding to the same DM configuration. In Figs. \ref{fig:4} (a) and \ref{fig:4} (b), the DM acts as a plane mirror. The auxiliary pump is sharply focused on the CCD, while the entangled photons exhibit strong correlations evidenced by the sharp peak in the center, as expected in the absence of any phase modulation. In Figs.~\ref{fig:4}(c) and \ref{fig:4} (d), the DM applies a phase corresponding to a linear combination of odd-parity Zernike polynomials. As predicted by theory, the auxiliary pump shows an interference pattern, but the two-photon correlations remain intact. The disorder introduced by the odd-parity phase has no effect on the spatial correlations of the entangled photons. Similarly, Figs.~\ref{fig:4} (e) and \ref{fig:4} (f) show the results for a different linear combination of odd-parity Zernike polynomials. Once again, the auxiliary pump experiences interference, while the two-photon correlations remain uncompromised. This confirms that the two-photon correlations are immune to odd-parity phase modulation, consistent with the theoretical prediction that only the even-parity component of the phase affects the entangled photons.

\begin{figure*}[t]
\includegraphics[width=\linewidth]{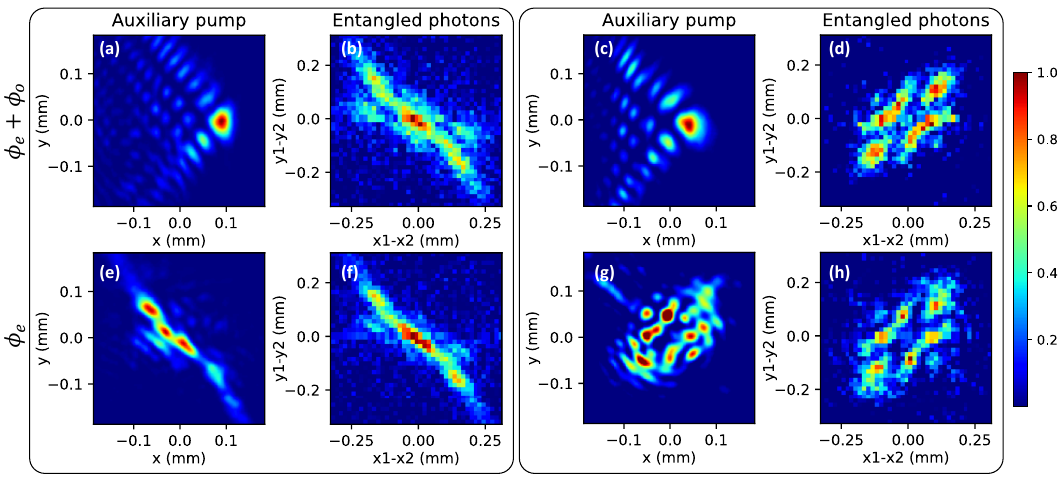}% Here is how to import EPS art
\caption{\label{fig:5} Experimental results for two different general disorders (first row) and their even-parity components (second row) implemented using DM. (a), (c), (e), and (g) show the auxiliary pump interference patterns, and (b), (d), (f), and (h) represent respective two-photon correlation in $x_1-x_2$ coincidence space.}
\end{figure*}

To further explore the impact of arbitrary phase disorder, we applied random linear combinations involving both odd and even Zernike polynomials to the deformable mirror (DM). Specifically, we added two different linear combinations of even-parity Zernike polynomials to the odd-parity configuration used in Fig.~\ref{fig:4} (e). The results are presented in Fig.~\ref{fig:5} (a)-(d).
In this case, no clear correlation is observed between the two-photon interference pattern and the auxiliary pump interference pattern. Next, we eliminated the odd-parity component by setting the coefficients of all odd Zernike polynomials to zero, leaving only the even-parity phase modulation. The corresponding results are shown in Fig.~\ref{fig:5}(e)-(h). Here, despite the absence of overall correlation in the total disorder case, a visible agreement emerges between the two-photon interference pattern and the auxiliary pump interference pattern. This confirms that the two-photon interference is exclusively affected by the even-parity component, while the odd-parity component has no influence. The same observation is re-endorsed by the agreement between (b) and (f), as well as (d) and (h). The slight mismatch between the two-photon and auxiliary pump interference patterns for the even-parity disorder can be attributed to two main factors:
(a) The finite Schmidt number of the entangled photons. As seen in Fig.~\ref{fig:4} (a), even without disorder, the two-photon correlation is not a Dirac delta function, leading to an averaging effect over nearby pixels. (b) Residual contributions from minor odd-parity Zernike modes due to experimental imperfections, which are ignored by the two-photon field. For these experiments, we acquired $200,000$ frames for the configurations in Fig.~\ref{fig:5} (b) and Fig.~\ref{fig:5} (f), and $300,000$ frames for the configurations in Fig.~\ref{fig:5} (d) and Fig.~\ref{fig:5} (h). 

While these experimental data convincingly support our theoretical claims, we extrapolate the claims into the regime of strong disorder. We employ suitable simulations to navigate the disorder regime outside of the hardware capabilities of the DM.

\begin{figure*}[t]
\includegraphics[width=\linewidth]{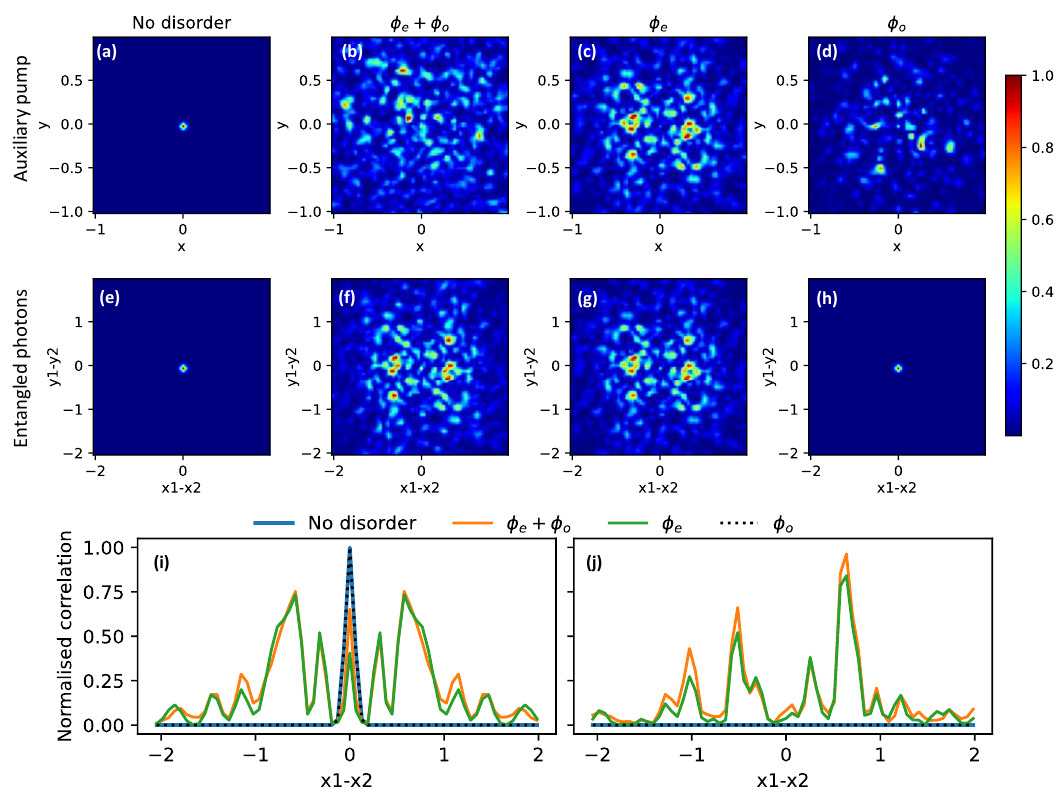}
\caption{\label{fig:6} Result of numerical simulation with a two-dimensional double Gaussian wavefunction in momentum variable illuminating a random phase mask and observed in the far field. (a)-(d) represent auxiliary pump intensity, and (e)-(h) represents two-photon correlation in coincidence space as a function of difference in detector positions for the same disorder configuration. (a), (e) are the results without any disorder. (b), (f) are the results when a general phase is added to the wavefronts. (c), (g) are the results corresponding to the even-parity part of the general phase, and (d), (h) are the results for the odd-parity part. (i) and (f) show the 1D cross-section of the two-photon correlation corresponding to the above disorder configurations along $y_1-y_2 = 0$ and $y_1-y_2 = 0.64$ for comparison.}
\end{figure*}

\section{Simulation}
We perform numerical simulations by considering a two-dimensional double Gaussian wavefunction in momentum space and a Gaussian auxiliary pump illuminating a random phase mask. Far field interference patterns are observed for both by applying a Fourier transform to the field after the disorder, followed by taking the squared modulus. To observe the two-photon interference, the 4-dimensional wavefunction is projected along $x_1 - x_2$ vs. $y_1 - y_2$, revealing the spatial correlations between photon pairs. The Schmidt number of this simulated two-dimensional wavefunction is $\approx1600$.
Fig.~\ref{fig:6} presents the results of numerical simulations. The first row of Fig.~\ref{fig:6} shows the auxiliary pump interference patterns and the second row presents the two-photon interference patterns corresponding to the same random phase configurations. Fig.~\ref{fig:6} (a) and ~\ref{fig:6} (e) correspond to the case with no disorder ($\phi(k) = 0$). As expected, the auxiliary pump shows focusing in the far field, while the two photons show perfect correlation, which manifests as a sharp peak in the two-photon interference pattern. When a random phase disorder is introduced, the auxiliary pump field gets modified and exhibits a strong speckle pattern shown in Fig.~\ref{fig:6} (b). Correspondingly, the correlation between the two photons is randomized, and a two-photon speckle pattern is observed in Fig.~\ref{fig:6} (f). Given that the disorder affects the two systems differently, there is no visible pattern correlation between the two-photon speckle and the auxiliary pump speckle.

In the scenario of the even-parity component of the same disorder (Fig.~\ref{fig:6} (c) and (g)), the auxiliary pump speckle pattern and the two-photon speckle pattern exhibit a strong visual match. The correlation coefficient between the two images is calculated to be $0.99$. For the record, the correlation coefficient between (f) and (g) is $0.94$. The slight deviation from $1$ is confirmed to arise from the finite Schmidt number of any SPDC process. For the odd-parity part of disorder, Fig.~\ref{fig:6} (d) shows the auxiliary pump speckle. However, the two-photon speckle vanishes and a single bright spot at the center appears, endorsing strong two-photon correlation, as seen in Fig.~\ref{fig:6} (h). This confirms the immunity of two-photon interference to the odd parity disorder. In the interest of better visual clarity, Fig.~\ref{fig:6} (i) and Fig.~\ref{fig:6} (j) present the one-dimensional cross-sections of the above data. The cross-section along $y_1 - y_2 = 0$ and $y_1 - y_2 = 0.64$ from Fig.~\ref{fig:6} (e)-(h) are plotted in Fig.~\ref{fig:6} (i) and (j), respectively. The cross-sections help in visualising the above conclusions lucidly.

The consequence of the finite Schmidt number can be understood as follows. Since the phase in k-space is expressed as the sum of odd and even parity terms ($\phi_o(k) + \phi_e(k)$), the corresponding disorder is a product of odd and even parity disorders ($\exp(i\phi_o(k))\cdot\exp(i\phi_e(k))$). In x-space, this implies that the field due to the disorder is the convolution of the fields from the even and odd parity components. In this convolution, the two-photon intensity caused by $\phi_o$ plays the part of a Dirac-$\delta$ function due to its sharply-peaked character (Fig.~\ref{fig:6} (h)). Consequently, the two-photon interference pattern for $\phi_e + \phi_o$ closely matches the pattern produced by $\phi_e$ alone (Fig.~\ref{fig:6} (i) and \ref{fig:6} (j). The slight discrepancy between the two patterns arises from the fact that the odd-parity two-photon intensity correlation (Fig.~\ref{fig:6} (h)) is only an approximate $\delta$-function, and has a finite width. This finite width is known to be a consequence of the finite Schmidt number of the SPDC process, as discussed in Reference \cite{Law2004}.

\section{Conclusion}
In conclusion, we investigated the effect of disorder on spatially entangled photons. We presented theoretical arguments to show that the two-photon correlation is not affected by odd-parity disorder. For an even-parity disorder, the two-photon interference pattern looks similar to an auxiliary pump interference. Although the addition of an odd-parity disorder to an even-parity one affects the auxiliary pump interference, it doesn't have any effect on the two-photon interference. Thus, our results show that two-photon correlations are partially immune to the disorder. We followed up with experiments using a deformable mirror (DM). Experimentally, the odd and even parities were generated using respective Zernike polynomials generated in the DM.  The experiments reproduced the theoretical predictions to an excellent degree. Subsequently, we followed up the experiments with numerical simulations that realized a strong disorder, and showed that our claims remain valid in the strong disorder regime as well.

The auxiliary pump interference pattern carries all the information about the disorder that is also invoked in two-photon interference pattern. Therefore, the auxiliary pump speckle, which is obviously stronger in intensity than the two-photon speckle, can be used as feedback for correcting the two-photon speckle. Our work clearly implies that wavefront correction using adaptive optics can be made faster since only the even-parity component of the aberration needs to be corrected. Consequently, this approach reduces the required number of independent optical elements for optimization by half. This will also reduce the correction time by a significant degree, given the smaller number of iterations needed in the wavefront correction process. This fact has clear implications in quantum imaging and quantum communication.

\begin{acknowledgments}
We acknowledge the TIFR CC HPC facility used for the numerical simulations. 
We express our gratitude to the Department of Atomic Energy, Government of India, for funding for Project Identification No. RTI4002 under DAE OM No. 1303/1/2020/R\&D-II/DAE/5567, Ministry of Science and Technology, India. The authors declare no conflict of interest.
\end{acknowledgments}

\section*{Data Availability Statement}

The data that support the findings of this study are available from the corresponding author upon reasonable request.

\appendix

\section{Deformable Mirror (DM) and Zernike polynomials}\label{ap:DM}
\begin{figure}[h!]
    \includegraphics[width=\linewidth]{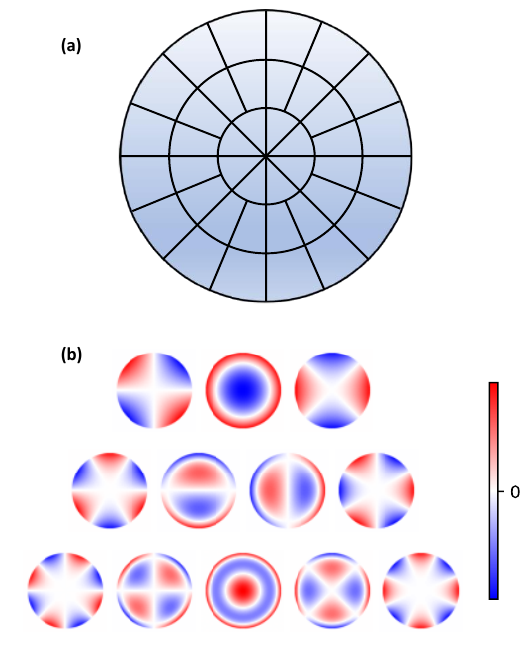}
    \caption{\label{fig:7} (a) The layout of the $40$ segments of the DM. (b) Structure of Zernike polynomials $\mathrm{Z_4}-\mathrm{Z_{15}}$. The first row shows $\mathrm{Z_4}$ to $\mathrm{Z_{6}}$ (left to right), which have even parity, the second row ($\mathrm{Z_7}$ to $\mathrm{Z_{10}}$) odd parity, and the third row ($\mathrm{Z_{11}}$ to $\mathrm{Z_{15}}$) also even parity.}
    \end{figure}

In this section, we describe the structure of the DM used in the experiment.
The DM features a continuous reflective surface with a total mirror diameter of $12.7$ mm and a pupil diameter of $10$ mm. The deformation of the mirror surface is controlled by $40$ independent piezoelectric actuators positioned on the back side of the reflective surface, as illustrated in Fig.~\ref{fig:7}(a). The center of each segment within the circular disc of the DM indicates the position of an actuator. Although these actuators can in principle be controlled independently, any change in the voltage of one actuator is observed to affect its boundary with neighboring segments compromising the independence of the individual pixels.

    \begin{table}[ht]
    \caption{\label{ztable}
    The first $15$ Zernike polynomials in polar coordinates, along with their parity, are listed below: }
    \begin{ruledtabular}
    \begin{tabular}{ccc}
    j & $\mathrm{Z_j}$ & Parity\\
    \hline
    1& 1 & Even\\
    
    2& $\rho \cos(\theta)$ & Odd\\
    3& $\rho \sin(\theta)$ & Odd\\
    
    4& $\sqrt{6}\rho^2\sin(2\theta)$& Even\\
    5& $\sqrt{3}(2\rho^2-1)$& Even\\
    6& $\sqrt{6}\rho^2\cos(2\theta)$& Even\\

    7& $\sqrt{8}\rho^3\sin(3\theta)$& Odd\\
    8& $\sqrt{8}(3\rho^3-2\rho)\sin(\theta)$& Odd\\
    9& $\sqrt{8}(3\rho^3-2\rho)\sin(\theta)$& Odd\\
    10& $\sqrt{8}\rho^3\cos(3\theta)$& Odd\\

    11& $\sqrt{10}r^4\sin(4\theta)$& Even\\
    12& $\sqrt{10}(4\rho^4-3\rho^2)\sin(2\theta)$ & Even\\
    13& $\sqrt{5}(6\rho^4-6\rho^2+1)$ & Even\\
    14& $\sqrt{10}(4\rho^4-3\rho^2) \cos(2\theta)$& Even\\
    15& $\sqrt{10}r^4\cos(4\theta)$ & Even\\
    
    \end{tabular}
    \end{ruledtabular}
    \end{table}

    On the other hand, the entire mirror surface can be employed to generate Zernike polynomials independently by suitably modulating the actuators. Any deformation on a continuous sheet of unit radius can be expressed as a linear combination of Zernike polynomials \cite{Zernike1934, Liu2022}. These polynomials form a complete basis for the deformations of the DM \cite{Smarra2022}. The DM, equipped with $40$ actuators and $3$ tilt controls, provides independent control over the first $15$ Zernike polynomials, which are listed in Table~\ref{ztable}. Zernike polynomials are a powerful and versatile mathematical tool for representing and correcting optical aberrations and atmospheric turbulence \cite{Noll:76}. Additionally, Zernike polynomials have well-defined parity (Table~\ref{ztable}), making them an ideal choice for our experiment. 
    
    The Zernike polynomials $\mathrm{Z_1}$, $\mathrm{Z_2}$ and $\mathrm{Z_3}$ represent the forward movement, horizontal tilt, and vertical tilt of the entire reflective surface, respectively, and can be controlled using the tilt controls. The shapes of the remaining Zernike polynomials ($\mathrm{Z_4}$ to $\mathrm{Z_{15}}$) are shown in Fig.~\ref{fig:7}(b). The first row depicts $\mathrm{Z_4}$ to $\mathrm{Z_{6}}$ (from left to right), which have even parity. The second row shows $\mathrm{Z_7}$ to $\mathrm{Z_{10}}$, which have odd parity, and the third row illustrates $\mathrm{Z_{11}}$ to $\mathrm{Z_{15}}$, which again have even parity. Thus, any random linear combination of odd Zernike polynomials results in an odd-parity disorder, while a linear combination of even Zernike polynomials creates an even-parity disorder. A random combination of all Zernike polynomials, regardless of parity, produces a random disorder without a defined parity. These deformations can be observed by sending a collimated light beam to illuminate the DM’s pupil and imaging the beam in the Fourier plane, which is achieved by placing a detector in the back focal plane of a lens. A perfectly flat mirror will generate a sharp, focused beam in the Fourier plane, while any deformations in the wavefront will introduce additional spatial frequencies, affecting the sharp focus.

\section{Measurement of Schmidt number}\label{ap:SNo}
    Here we discuss how we measure Schmidt number of the SPDC photons. The two-photon wavefunction in terms of transverse momentum can be approximated by double Gaussian written in Eq. \ref{eq:one}. The Schmidt number is given by \cite{Law2004}:
    \begin{eqnarray}
        K = \frac{1}{4}\left(\frac{\sigma_+}{\sigma_-} + \frac{\sigma_-}{\sigma_+}\right)^2 
        \label{eq:AB1}
    \end{eqnarray}
    There are two variables which we need to measure in order to obtain the value of Schmidt number. The two variables can be obtained from two measurements: the standard deviation of (a) joint probability distribution (JPD) of two photons (can be obtained by coincidence measurements), and (b) marginal probability distribution of one of the photons (given by the actual beam).

    The JPD of two-photons is given by:
    \begin{eqnarray}
        \rho(k_1,k_2) \propto e^{-(k_1+k_2)^2\sigma_+^2}e^{-(k_1-k_2)^2\sigma_-^2}
    \end{eqnarray}
    As $\sigma_+\gg \sigma_-$, the momenta of two-photons are anti-correlated. Hence, the projection of JPD in $k_1+k_2$ space will be sharply peaked and the width of anti-correlation can be used first measurement. The standard deviation of of JPD in $k_1+k_2$ space, $\sigma_j = \frac{1}{\sqrt{2}\sigma_+}$. 
    
    \begin{figure}
    \includegraphics[width=\linewidth]{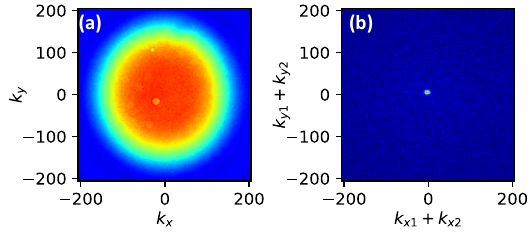}
    \caption{\label{fig:8} 
        (a) The original beam recorded using EMCCD camera in the far field. (b) Two-photon momentum anti-correlation.}
    \end{figure}
    
    The marginal distribution of one of the photons can be obtained by integrating over the other photon in the JPD and is given by:
    \begin{eqnarray}
        \rho(k_1) \propto e^{-k_1^2 \frac{4\sigma_+^2\sigma_-^2}{\sigma_+^2+\sigma_-^2}}
    \end{eqnarray}
    For high Schmidt number with $\sigma_+\gg \sigma_-$, the standard deviation of the marginal probability distribution will be, $\sigma_m = \frac{1}{{2\sqrt{2}\sigma_-}}$. 
    The square of  the ratio of the $\sigma_m$ and $\sigma_j$ will be:
    \begin{eqnarray}
        \left(\frac{\sigma_m}{\sigma_j}\right)^2 = \frac{1}{4}\left(\frac{\sigma_+}{\sigma_-}\right)^2
    \end{eqnarray}
    which is equal to the Schmidt number defined in Eq. \ref{eq:AB1} for $\sigma_+\gg \sigma_-$. Hence, the Schmidt number can be calculated using $\sigma_m$ and $\sigma_k$.

    To experimentally obtain $\sigma_m$ and $\sigma_j$, the measurements need to be performed in the far field (momentum space), which can be obtained by inserting a lens after the nonlinear crystal and detecting the photons in the back focal plane. Fig.~\ref{fig:8} (a) shows the beam captured using an EMCCD which gives the marginal distribution of a photon and Fig.~\ref{fig:8} (b) shows the projection of JPD in $k_1+k_2$ space  obtained by taking $3000$ frames using EMCCD and post-analyzing using the method discussed in Reference \cite{MTh}. The values of $\sigma_m$ and $\sigma_j$ obtained by 2D Gaussian fitting are $121.97$ and $4.59$ pixels respectively. Therefore the Schmidt number of SPDC photons is $707\pm10$.

\section{Comparing the interference pattern for two wavelengths}\label{ap:SC}
\begin{figure}
    \includegraphics[width=\linewidth]{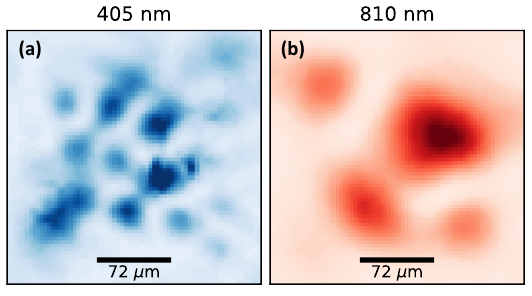}
    \caption{\label{fig:9} Interference pattern observed with coherent light source of wavelength (a) $405$ nm and (b) $810$ nm for same configuration of the DM.}
    \end{figure}

    Here, we compare the speckle patterns corresponding to two wavelengths, $405$ nm and $810$ nm, for identical configurations of the DM. A collimated coherent light source with a waist size comparable to the DM's radius is used to illuminate the DM at both wavelengths, and the interference patterns are observed in the far field. The results are shown in Fig.~\ref{fig:9}.
    Due to the wavelength dependence of the phase for a given path length, the $405$ nm light experiences double the phase shift compared to the $810$ nm light. Consequently, the speckle patterns for the two wavelengths are not expected match. The strength of disorder introduced by the DM can be quantified using the speckle contrast, defined as:
    \begin{eqnarray}
        \kappa = \frac{\sigma_I}{\langle I\rangle}
    \end{eqnarray}
    where $\sigma_I$ represents the standard deviation of the intensity distribution, and $\langle I\rangle$ denotes the mean intensity. This measure provides a quantitative comparison of the disorder for the two wavelengths.
    The speckle contrast values for the two wavelengths ($405$ nm and $810$ nm) are $0.89$ and $0.96$ respectively, indicating that the disorder strength is comparable for both.

\nocite{*}
\bibliography{aipsamp}

\end{document}